\documentclass[preprint,doublespace,showpacs]{revtex4} 

\usepackage{graphicx}

\usepackage{dcolumn}
\usepackage{bm}
\usepackage{amssymb}
\usepackage{amsmath}

\begin{document}

\title{Intermediate-field two-photon absorption enhancement
by shaped femtosecond pulses with spectral phases of antisymmetric nature}

\author{Lev Chuntonov, Leonid Rybak, Andrey Gandman, and Zohar Amitay}
\email{amitayz@tx.technion.ac.il} %
\affiliation{Schulich Faculty of Chemistry, Technion - Israel
Institute of Technology, Haifa 32000, Israel}

\begin{abstract}
We demonstrate and study the enhancement of intermediate-field two-photon absorption
by shaped femtosecond pulses having spectral phases of antisymmetric nature.
%
%
The intermediate-field regime corresponds to pulse intensities,
where the two-photon absorption is coherently induced
by the weak-field nonresonant two-photon transitions as well as
by additional resonance-mediated four-photon transitions.
%
It is a regime of significant excitation yields, 
exceeding the weak-field yields by two orders of magnitudes,
reaching about 10-20$\%$ population transfer.
%
The considered antisymmetric nature 
is with respect to one-half
of the (initial-to-final) two-photon transition frequency.
The corresponding pulse spectrum is detuned from this frequency
(the detuning direction is according to the system).
We study in detail
the coherent interference mechanism leading to the observed 
enhancement using forth-order frequency-domain perturbative analysis.
We also show that,
even though the maximal 
enhancement is achieved with phase patterns of perfect antisymmetry,
at high enough intermediate-field intensities
absorption enhancement beyond the transform-limited level
is still achievable even with 
patterns having 
some degree of deviation from perfect antisymmetry. 
The 
degree of tolerance to deviations from perfect antisymmetry
increases as the pulse intensity increases.
The theoretical and experimental model system of the study is atomic sodium. 
These findings are of particular importance for coherent control scenarios
that simultaneously involve multiple excitation channels.
\end{abstract}



\maketitle


\section{Introduction}

Coherent control of multiphoton processes in atoms and molecules
using shaped femtosecond pulses is a subject of both fundamental and
applicative importance \cite{tannor_kosloff_rice_coh_cont,
shapiro_brumer_coh_cont_book, warren_rabitz_coh_cont,
rabitz_vivie_motzkus_kompa_coh_cont, gordon_rice_coh_cont,
gerber_feedback_control_review,
dantus_exp_review1_2,silberberg_2ph_nonres1_2,
dantus_2ph_nonres_molec1_2, baumert_2ph_nonres,
silberberg_2ph_1plus1, girard_2ph_1plus1, becker_2ph_1plus1_theo,
amitay_multi-channel, whitaker_3ph_iodine_chirp,
silberberg_antiStokes_Raman_spect,
gersh_murnane_kapteyn_Raman_spect, leone_res_nonres_raman_control,
leone_cars, motkus_cars, silberberg-2ph-strong-field,
silberberg_2D_maps_chirp, weinacht-2ph-strong,
wollenhaupt-baumert1_2, amitay_3ph_2plus1,
amitay_2ph_inter_field1,amitay_2ph_inter_field2}.
The broad spectrum of the femtosecond pulses photo-induces a coherent manifold
of state-to-state initial-to-final pathways that by manipulating
their interference state-to-state transition probabilities are controlled.
Inducing constructive or destructive interferences among these pathways
leads to either an enhancement or attenuation, respectively,
of the transition probability.
It is achieved by tailoring the shape of the femtosecond pulse.
Experimentally, it is implemented using pulse shaping techniques
that are applied in the frequency domain to control the spectral
characteristics of the pulse \cite{pulse_shaping}.

Rational design of the pulse shape is based on identifying the
state-to-state interfering pathways and their interfering
mechanisms.
As the pulse shaping is conducted in the frequency domain, rational
pulse design is most powerful when the potoexcitation picture is
available in the frequency domain.
This is possible once the excitation is validly described by the
time-dependent perturbation theory of finite order, following a
proper Fourier transformation to the frequency domain.
Such rational control has been demonstrated successfully in many
cases \cite{dantus_exp_review1_2, silberberg_2ph_nonres1_2,
dantus_2ph_nonres_molec1_2, baumert_2ph_nonres,
silberberg_2ph_1plus1, girard_2ph_1plus1, becker_2ph_1plus1_theo,
amitay_multi-channel, whitaker_3ph_iodine_chirp,
silberberg_antiStokes_Raman_spect,
gersh_murnane_kapteyn_Raman_spect, leone_res_nonres_raman_control,
leone_cars, motkus_cars, silberberg-2ph-strong-field,
silberberg_2D_maps_chirp, weinacht-2ph-strong,
wollenhaupt-baumert1_2, amitay_3ph_2plus1,
amitay_2ph_inter_field1,amitay_2ph_inter_field2}.
However, until recently, it has been implemented 
only in the weak-field regime \cite{dantus_exp_review1_2,
silberberg_2ph_nonres1_2, dantus_2ph_nonres_molec1_2,
baumert_2ph_nonres, silberberg_2ph_1plus1, girard_2ph_1plus1,
becker_2ph_1plus1_theo,  amitay_multi-channel,
whitaker_3ph_iodine_chirp, silberberg_antiStokes_Raman_spect,
gersh_murnane_kapteyn_Raman_spect, leone_res_nonres_raman_control,
leone_cars, motkus_cars, amitay_3ph_2plus1}, where the multiphoton
process is described by the time-dependent perturbation theory of
the lowest nonvanishing order.
For N-photon process it is the Nth perturbative order.
%
The simplicity of this theoretical description is the attractiveness of the weak-field regime.
Still, it 
has a considerable downside of being limited
to low excitation yields, typically up to only $\sim$0.1$\%$ population transfer.
%
%
%
%
Recently, we have extended theoretically and experimentally
the rational pulse design approach 
to the intermediate-field regime
\cite{amitay_2ph_inter_field1,amitay_2ph_inter_field2},
where the excitation yields exceed by two orders of magnitude the weak-field yields,
reaching 10-20$\%$ population transfer.
%
In this regime, the perturbative description of the multiphoton
process needs to additionally include also the next non-vanishing order beyond the
lowest one, i.e. orders N and N+2 for N-photon process.

Specifically, we have studied the process of atomic two-photon
absorption \cite{amitay_2ph_inter_field1,amitay_2ph_inter_field2}.
In the weak-field regime it is described by second-order perturbation
theory, while in the intermediate-field regime it is described by
forth-order perturbation theory considering both the second and forth
perturbative orders.
Physically, the corresponding weak-field absorption involves
initial-to-final nonresonant excitation pathways that are composed of two
absorbed photons, while the intermediate-field absorption also
involves additional resonance-mediated pathways that are composed of three absorbed
photons and one emitted photon.
Consequently, as already has been shown a decade ago \cite{silberberg_2ph_nonres1_2},
the weak-field two-photo absorption is maximized
by the transform-limited (TL) having zero (relative) phase across the whole spectrum
as well as by any shaped pulse having a spectral phase that is antisymmetric around
one-half of the two-photon transition frequency $\omega_{fg}$.
On the other hand, as we have recently shown in short \cite{amitay_2ph_inter_field1},
in the intermediate-field regime
the shaped pulses with antisymmetric phase patterns actually enhance the absorption beyond the 
level induced by the TL pulse. It occurs when the center of the pulse spectrum is
detuned from $\omega_{fg}/2$, either to the red or to the blue according to the system.

The enhancement of the intermediate-field two-photon absorption by shaped
pulses with spectral phase patterns of antisymmetric nature is the subject of the present
theoretical and experimental work.
The coherent interference mechanism leading to the observed enhancement is
studied and analyzed in detail using forth-order frequency-domain perturbative analysis.
We also show that,
even though the maximal enhancement is achieved with phase patterns of perfect antisymmetry,
at high-enough intermediate-field intensities
absorption enhancement beyond the TL level
is still achievable even with 
patterns having 
some degree of deviation from perfect antisymmetry. 
The corresponding
degree of tolerance to deviations from perfect antisymmetry
increases as the pulse intensity increases.
The theoretical and experimental model system of the study is atomic sodium (Na).
Beyond the basic scientific understanding,
these findings are also of significance for coherent control scenarios
that simultaneously involve multiple excitation channels.
For example, the tolerance to deviation from perfect antisymmetry
allows to keep the excitation of the two-photon channel sufficiently high,
while adjusting the excitation of the other channels.

The paper is organized as follows.
Section~II gives 
first the frequency-domain theoretical description of the
intermediate-field femtosecond two-photon absorption process and then refers 
to the case when the absorption is induced by shaped pulses
having phase patterns of antisymmetric nature.
%
Section~III presents the corresponding experimental and theoretical-numerical results for
the intermediate-field two-photon absorption in Na,
and Sec.~IV analyzes and discusses them in detail.
The paper ends with conclusions in~Sec.~V.

\section{Theoretical description}
\subsection{Intermediate-field description of femtosecond two-photon absorption}

\begin{figure} 
\includegraphics[width=17cm]{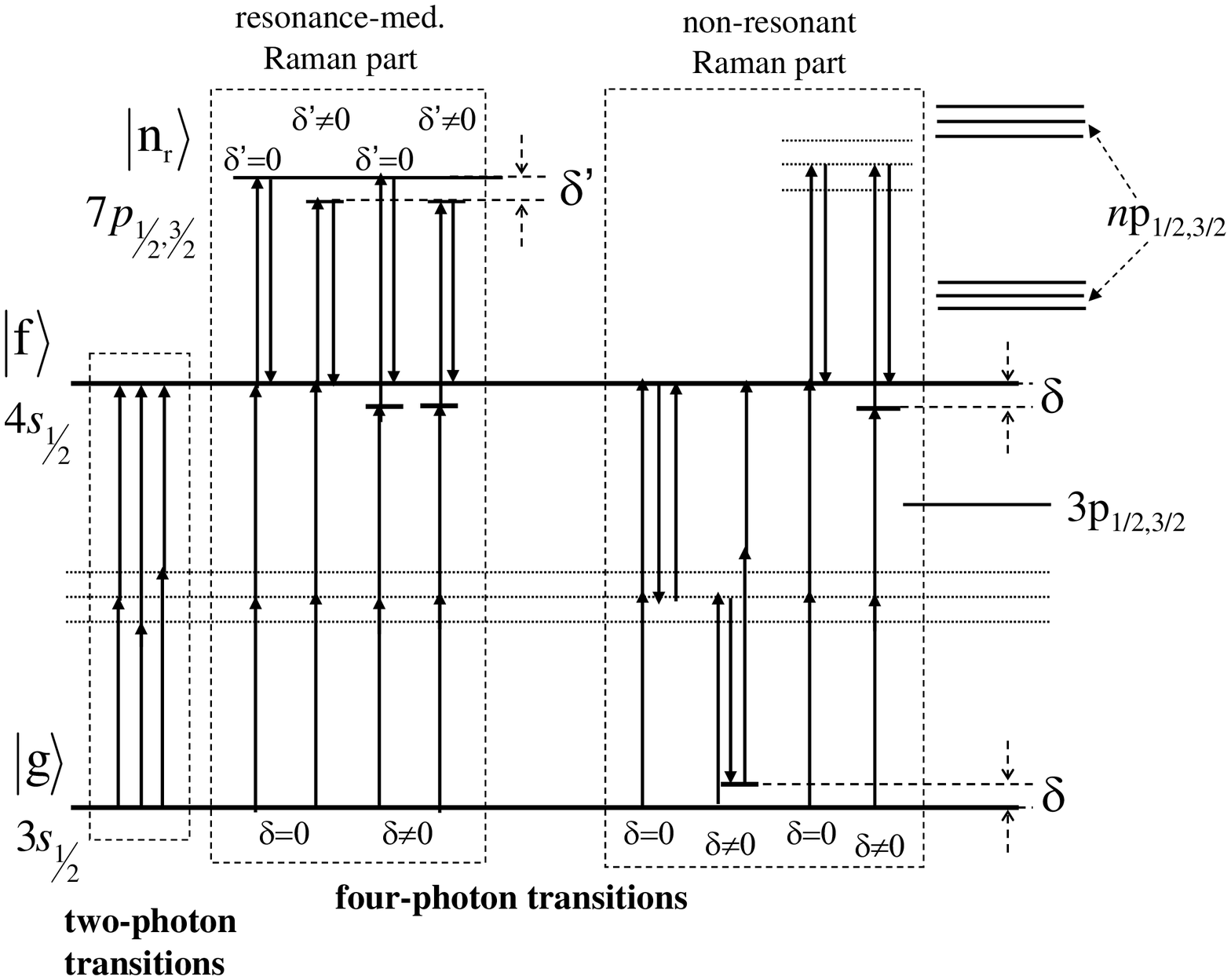}     
\vspace*{-0.7cm}
\caption{Excitation scheme of femtosecond
two-photon absorption in the intermediate-field regime. The
indicated levels correspond to the Na atom (not to scale). Shown are
pathway examples of non-resonant two-photon transitions and
four-photon transitions
from $\left| f \right\rangle \equiv 4s$ to $\left| g \right\rangle \equiv 3s$. 
The four-photon transitions involve three absorbed photons and one
emitted photon in any possible order, and thus can be decomposed
into two parts: a non-resonant two-photon transition and a Raman
transition.
The border line between these two parts can be either on-resonance
or near-resonance with $3s$ or $4s$ (with detuning $\delta$). The
Raman transition itself can be non-resonant due to the $np$ states
(except for $7p$) or on/near-resonance with $7p$ (with detuning
$\delta'$).
} \label{fig_1}
\end{figure}

The atomic femtosecond two-photon absorption under consideration is
from a ground state $\left|g\right>$ to an excited state
$\left|f\right>$, which are coupled via a manifold of states
$\left|n\right>$ having the proper symmetry.
The pulse spectrum is such that all the
$\left|g\right>$-$\left|n\right>$ and $\left|f\right>$-$\left|n\right>$ couplings are non-resonant,
except for the $\left|f\right>$-$\left|n_{r}\right>$ resonant coupling.
In other words, the spectral amplitude is zero at all the
$\left|g\right>$-$\left|n\right>$ and $\left|f\right>$-$\left|n\right>$ transition frequencies,
i.e., $\left|E(\omega_{gn})\right| = \left|E(\omega_{fn})\right| = 0$,
except for the $\left|f\right>$-$\left|n_{r}\right>$ transition frequency,
i.e., $\left|E(\omega_{fn_{r}})\right| \ne 0$.
The corresponding excitation scheme is shown in Fig.~\ref{fig_1}.

Within the present intermediate-field regime, the time-dependent
(complex) amplitude $A_{f}$ of state $\left|f\right>$, following
irradiation with a (shaped) temporal electric field
$\varepsilon(t)$, can be validly described by 4$^{th}$-order
time-dependent perturbation theory. Generally, it includes
non-vanishing contributions from both the 2$^{nd}$ and 4$^{th}$
perturbative orders:
\begin{equation} A_{f} =
A_{f}^{(2)} + A_{f}^{(4)}.   \label{eq4:time-amp}
\end{equation}
The perturbative description allows a transformation into a
frequency-domain, where the spectral field of the pulse $E(\omega)
\equiv \left|E(\omega)\right| \exp \left[ i\Phi(\omega) \right]$ is
given as the Fourier transform of $\varepsilon(t)$, with
$\left|E(\omega)\right|$ and $\Phi(\omega)$ being, respectively, the
spectral amplitude and phase of frequency $\omega$.
For the unshaped transform-limited (TL) pulse, $\Phi(\omega)=0$ for
any $\omega$.
We also define the normalized spectral field $\widetilde{E}(\omega)
\equiv E(\omega) / \left|E_{0}\right| \equiv
\left|\widetilde{E}(\omega)\right| \exp \left[ i\Phi(\omega)
\right]$ that represents the pulse shape, where $\left|E_{0}\right|$
is the peak spectral amplitude.
This allows to clearly distinguish in the expressions given below
between the dependence on the pulse intensity and the dependence on
the pulse shape.
The maximal spectral intensity $I_{0}$ is proportional to
$|E_{0}|^{2}$ ($I_{0}\propto|E_{0}|^{2}$) and correspond to
different temporal peak intensities $I_{TL}$ of the
transform-limited (TL) pulse.

As shown before for the weak-field regime
\cite{silberberg_2ph_nonres1_2}, the 2$^{nd}$-order amplitude
$A_{f}^{(2)}$ is given by
\begin{eqnarray}
A_{f}^{(2)} & = & - \frac{1}{i \hbar^{2}} \left|E_{0}\right|^{2}
A^{(2)}(\omega_{fg}) \; ,
\label{eq5:tot-amp-2nd-order} \\
A^{(2)}(\Omega) & = & \mu_{fg}^{2}
\int_{-\infty}^{\infty}\widetilde{E}(\omega)\widetilde{E}(\Omega-\omega)d\omega
\; , \label{eq6:expl-amp-2nd-order-w_fg}
\end{eqnarray}
where $\omega_{fg}$ is the $\left|g\right>$-$\left|f\right>$
transition frequency and $\mu_{fg}^{2}$ is the corresponding real
effective non-resonant two-photon coupling.
Equations~(\ref{eq5:tot-amp-2nd-order}) and (\ref{eq6:expl-amp-2nd-order-w_fg})
reflect the fact that $A_{f}^{(2)}$ coherently interferes all the
non-resonant two-photon pathways from $\left|g\right>$ to
$\left|f\right>$ of any combination of two absorbed photons with
frequencies $\omega$ and $\omega' = \omega_{fg} - \omega$,
i.e., having their frequency sum equal to $\omega_{fg}$.
Several such two-photon pathways are shown schematically in Fig.~\ref{fig_1}.
%

The 4$^{th}$-order amplitude term $A_{f}^{(4)}$ is given by
\begin{eqnarray}
A_{f}^{(4)} & = & - \frac{1}{i \hbar^{4}} \left|E_{0}\right|^{4}
\left[ A_{f}^{(on-res)} + A_{f}^{(near-res)} \right] \; ,
\label{eq7.1:amp-4th-order} \\
A_{f}^{(on-res)} & = & i \pi A^{(2)}(\omega_{fg}) A^{(R)}(0) \; ,
\label{eq7.2:amp-4th-order} \\
A_{f}^{(near-res)} & = & - \wp \int_{-\infty}^{\infty} d\delta
\frac{1}{\delta} A^{(2)}(\omega_{fg}-\delta) A^{(R)}(\delta) \; ,
\label{eq7.3:amp-4th-order}
\end{eqnarray}
where $A^{(2)}(\Omega)$ is defined in
Eq.~(\ref{eq6:expl-amp-2nd-order-w_fg}) and
\begin{eqnarray}
A^{(R)}(\Delta\Omega) & = & A^{(non-resR)}(\Delta\Omega) +
A^{(resR)}(\Delta\Omega) \; ,
\label{eq8:amp-4th-order}\\
A^{(non-resR)}(\Delta\Omega) & = & (\mu_{ff}^{2}+\mu_{gg}^{2})
\int_{-\infty}^{\infty} \widetilde{E}(\omega + \Delta\Omega)
\widetilde{E}^{*}(\omega) d\omega  \; ,
\label{eq9:amp-4th-order} \\
A^{(resR)}(\Delta\Omega) & = & |\mu_{fn_{r}}|^{2} \left[ i \pi
\widetilde{E}(\omega_{fn_{r}}+\Delta\Omega)
\widetilde{E}^{*}(\omega_{fn_{r}}) \right. \hspace{5cm} \nonumber \\
& & \left.
         - \wp \int_{-\infty}^{+\infty} d\delta' \frac{1}{\delta'} \widetilde{E}(\omega_{fn_{r}}+\Delta\Omega-\delta')
                \widetilde{E}^{*}(\omega_{fn_{r}}-\delta') \right]. \hspace{0.7cm}
\label{eq10:amp-4th-order}
\end{eqnarray}
%
This set of equations reflects the fact that $A_{f}^{(4)}$
interferes all the four-photon pathways from $\left|g\right>$ to
$\left|f\right>$ of any combination of three absorbed photons and
one emitted photon. Several typical four-photon pathways are shown
schematically in Fig.~\ref{fig_1}.

As seen in Fig.~\ref{fig_1} and reflected in the above equations,
each four-photon pathway is either on-resonant or near-resonant with
either $\left|g\right>$ or $\left|f\right>$ with a corresponding
detuning of $\delta$.
Accordingly, $A_{f}^{(4)}$ [Eq.(\ref{eq7.1:amp-4th-order})] has two contributions:
$A_{f}^{(on-res)}$ [Eq.(\ref{eq7.2:amp-4th-order})]
coherently interferes all the on-resonant pathways having $\delta=0$,
while $A_{f}^{(near-res)}$ [Eq.(\ref{eq7.3:amp-4th-order})]
coherently interferes all the near-resonant pathways having $\delta\ne0$.
The on-resonant pathways are excluded from the integration in
$A_{f}^{(near-res)}$ by the Cauchy's principle value operator $\wp$.
Each of the amplitudes $A_{f}^{(on-res)}$ and $A_{f}^{(near-res)}$
is divided into a sequence of two two-photon parts:
(i) a part that interferes all the non-resonant two-photon transitions with a frequency sum of $\Omega=\omega_{fg}-\delta$
and (ii) a part that interferes all the Raman transitions with a frequency difference of $\Delta\Omega=\delta$.
The border line between these two parts is detuned by $\delta$ from
$\left|f\right>$ or $\left|g\right>$ according to whether,
respectively, the two-photon part [(i)] or the Raman part [(ii)] comes first.
These two parts are expressed, respectively, by the parameterized amplitudes
$A^{(2)}(\Omega)$ [Eq.(\ref{eq6:expl-amp-2nd-order-w_fg})] and $A^{(R)}(\Delta\Omega)$ [Eq.(\ref{eq8:amp-4th-order})].

The Raman transitions are either of non-resonant nature
or of resonance-mediated nature via $\left|n_{r}\right>$.
The non-resonant Raman transitions are interfered in the
parameterized amplitude $A^{(non-resR)}$
[Eq.\ref{eq9:amp-4th-order}], with $\mu^{2}_{gg}$ and $\mu^{2}_{ff}$
being the $\left|g\right>$-$\left|g\right>$ and
$\left|f\right>$-$\left|f\right>$ real effective non-resonant Raman
couplings due to all the non-resonantly coupled states
$\left|n\right>$.
The resonance-mediated Raman transitions are interfered in the
parameterized amplitude $A^{(resR)}$ [Eq.\ref{eq10:amp-4th-order}],
with $\mu_{fn_{r}}$ being the $\left|f\right>$-$\left|n_{r}\right>$
dipole-moment matrix element.
The corresponding on-resonant $(\delta'=0)$ and near-resonant $(\delta'\ne0)$ pathways are separately interfered there,
with $\delta'$ being the detuning from $\left|n_{r}\right>$ (see Fig.~\ref{fig_1}).

For a given physical system and a given pulse shape
$\widetilde{E}(\omega)$, a non-zero $A_{f}^{(2)}$ is proportional to
$I_{0}$ while a non-zero $A_{f}^{(4)}$ is proportional to
$I_{0}^{2}$, i.e., their ratio is proportional to $I_{0}$.
For a given $I_{0}$, the relative magnitude and relative sign
between $A_{f}^{(2)}$ and $A_{f}^{(4)}$ are generally determined by
the pulse shape $\widetilde{E}(\omega)$ and by the magnitudes and
signs of the different Raman couplings ($\mu_{gg}^{2}$,
$\mu_{ff}^{2}$, and $|\mu_{fn_{r}}|^{2}$).
In the present work, for a set of intensities $I_{0}$, the final
$\left|f\right>$ population $P_{f}=\left|A_{f}\right|^{2}=\left|A_{f}^{(2)}+A_{f}^{(4)}\right|^2$
(i.e., the degree of the two photon absorption) is controlled via the pulse shape
$\widetilde{E}(\omega)$.

\subsection{Two-photon absorption by shaped pulses with antisymmetric phase patterns}

The weak-field two-photon absorption is associated with the
amplitude $A^{(2)}(\omega_{fg})$ [Eqs.(\ref{eq5:tot-amp-2nd-order})
and (\ref{eq6:expl-amp-2nd-order-w_fg})]. The phase associated with
each $\left|g\right>$-$\left|f\right>$ two-photon pathway is
$\Phi_{pathway}=\Phi(\omega) + \Phi(\omega_{fg}-\omega)$.
So, with the TL pulse all these pathways acquire zero phase $\Phi_{pathway}=0$
and thus interfere one with the other in a fully constructive way. For a
given spectrum $\left|E(\omega)\right|$, this leads to the maximal value of
$\left|A^{(2)}_{f}\right|$ and to the maximal degree of weak-field non-resonant
two-photon absorption \cite{silberberg_2ph_nonres1_2}.
Additionally \cite{silberberg_2ph_nonres1_2},
this maximal value of $\left|A^{(2)}_{f}\right|$ is also induced
by any shaped pulse having a spectral phase pattern $\Phi(\omega)$
that is antisymmetric around one-half of the two-photon transition frequency $\omega_{fg}/2$,
i.e., satisfying the relation $\Phi(\omega) = -\Phi(\omega_{fg}-\omega)$,
which also yields zero phase $\Phi_{pathway}=0$ for all the two-photon pathways.

We consider the phase patterns $\Phi(\omega)$ to be divided into 2$n$ bins,
each having a spectral width of $\Delta_{bin}$. 
The inversion point of the antisymmetric pattern is located between
the bins 2n/2 and 2n/2+1 and corresponds exactly to $\omega_{fg}/2$.
So, each antisymmetric pattern contains $n$ bin pairs such that
$\Phi_{m}(\omega_{fg}/2+\xi)=-\Phi_{m}(\omega_{fg}/2-\xi)$ for any
$\xi$ satisfying $\left\{(m-1)\Delta_{bin} \le  \xi < m \Delta_{bin}
\right\}$ with $m=1 \ldots n$.
If a bin does not have the same phase value as its pair partner,
the phase of the corresponding pathway is non-zero ($\Phi_{pathway} \ne 0$)
and, hence, this pathway does not interfere constructively with the other pathways.
This leads to the reduction of the two-photon absorption amplitude
$\left|A^{(2)}(\omega_{fg})\right|$ below its maximal value.
The larger the number of such non-matched bins is, the smaller is
the degree of two-photon absorption.

In the intermediate-field regime the final amplitude is contributed
by both the second- and fourth-order terms.
In the fourth-order term, the value of $A^{(2)}(\Omega=\omega_{fg})$
determines the contribution of the pathways that are on resonant
with $\left|g\right>$ or $\left|f\right>$
[Eq.(\ref{eq7.2:amp-4th-order})].
The actual shape of $A^{(2)}(\omega_{fg}\pm\delta)$ is also
important: following the integration in
Eq.(\ref{eq7.3:amp-4th-order}) it determines the contribution of the
pathways that are near resonant with $\left|g\right>$ or $\left|f\right>$.
In this work the spectrum of the pulse is chosen such that the
central spectral frequency $\omega_{0}$ is of red detuning from
$\omega_{fg}/2$, with $\left|\widetilde{E}(\omega_{fg}/2)\right| \simeq 0.5$.
Such a spectrum generally corresponds to the typical case, where
$A^{(4)}$ is negligible relative to $A^{(2)}$ in the weak-field
limit and becomes comparable to $A^{(2)}$ in the upper intermediate-field limit.

We have previously shown \cite{amitay_2ph_inter_field1} that,
with a pulse spectrum that is detuned to the red from one-half of the two-photon transition frequency,
the intermediate-field femtosecond two-photon absorption in atomic sodium
is enhanced beyond the TL level by shaped pulses having the antisymmetric phase patterns.
As in the weak-field regime, such patterns maximize the magnitude of the second-order amplitude
$\left|A^{(2)}(\Omega=\omega_{fg})\right|$ to the level of the TL pulse.
On the other hand, they lead to a fourth-order term with a much smaller magnitude
$\left|A^{(4)}(\Omega=\omega_{fg})\right|$ as compared to the TL pulse.
However, the inter-term interference between the second- and the fourth-order amplitudes is
actually of destructive nature.
Hence, the antisymmetric phase patterns suppress this destructive interference
and lead to an enhanced absorption as compared to the TL absorption.
In the work below we demonstrate that the requirement for perfect
antisymmetry can be relaxed to some degree.
Even if the phase pattern contains several bin pairs that their bin
members do not have the same phase magnitude and opposite sign, at
sufficiently high intensities (still within the intermediate-field
regime) such shaped pulses still enhance the two-photon absorption
beyond the TL level.
As explained below, this effect results from the destructive interference between the
second- and fourth-order terms, which is stronger for the TL pulse,
in combination with fact that higher intensity leads to stronger destructiveness as the
ratio between the fourth- and second-order terms increases with intensity (see above).
Also plays a role here the interference between the on- and
near-resonant components of the fourth-order amplitude, as they are
affected differently by the deviation from perfect antisymmetry of
the phase pattern.

\section{Results}
Our model system is atomic sodium \cite{NIST} with the excitation
diagram shown on Fig.~\ref{fig_1}.
The ground state $\left|g\right>$ corresponds to the $3s$ state,
the excited state $\left|f\right>$ corresponds to the $4s$ state,
and the $\left|n\right>$-states correspond to the various
$p$-states, with $\left|n_r\right>$ being the $7p$ state.
The two-photon transition frequency
$\omega_{f,g}\equiv\omega_{4s,3s}$=25740~cm$^{-1}$ corresponds to
two photons of 777~nm and the transition frequency
$\omega_{n_r,f}\equiv\omega_{7p,4s}$=12800~cm$^{-1}$ corresponds to
one photon of 781.2~nm.
The corresponding inversion point of the antisymmetric phase patterns is
the frequency $\omega_{f,g}/2\equiv\omega_{4s,3s}$/2=12870~cm$^{-1}$.

Experimentally, vapor of atomic sodium is produced in a static vacuum chamber at
$350^{\circ}$C 
with 10-Torr Ar buffer gas.
It is irradiated at a 1-kHz repetition rate with amplified
phase-shaped linearly-polarized femtosecond Gaussian laser pulses.
The central spectral wavelengths is $\lambda_{0}$=780~nm, the
corresponding bandwidth (FWHM) is 5~nm, and the TL pulse duration is
$\sim$180~fs.
The laser pulses undergo shaping in a 4$f$ optical setup
incorporating a pixelated liquid-crystal spatial light phase
modulator in the Fourier plane of the shaping setup \cite{pulse_shaping}.
%
%
The experiment is conducted with different pulse energies. Upon
focusing, the corresponding temporal peak intensity of the
transform-limited (TL) pulse at the peak of the spatial beam profile
$I_{TL}^{(profile-peak)}$ ranges from 5$\times$10$^{8}$ to
7$\times$10$^{10}$~W/cm$^2$.
Following the interaction with a pulse, the Na population excited to
the $4s$ state radiatively decays to the lower $3p$ state. The
fluorescence emitted in the decay of the $3p$ state to the $3s$
ground state serves as the relative measure for the excited $4s$
population $P_{f} \equiv P_{4s}$.
It is optically measured at 90$^{\circ}$ to the laser beam
propagation direction using a spectrometer coupled to a time-gated
camera system. The measured signal results from an integration over
the spatial beam profile.

We have divided the pulse spectral region around one-half of the
two-photon transition frequency $\omega_{4s,3s}/2$ into 2$n$=24
bins, with each bin having a width of $\Delta_{bin}$=6.15cm$^{-1}$.
The phase values of the $n$=12 left-hand bins have been randomized 
and then inverted and assigned to the right-hand bins to obtain the
antisymmetric pattern.
Our findings are demonstrated by measuring the two-photon absorption
induced by 2500 shaped pulses with different random anti-symmetric phase patterns
at pulse intensities ranging from the weak-field regime
to the upper limit of the intermediate-field regime.
The results are presented as histograms (distributions) of the
ratio between the two-photon absorption induced by the shaped pulse
and the two-absorption induced by the TL pulse.
We refer to this ratio as the enhancement factor.

\begin{figure} 
\includegraphics[width=17cm]{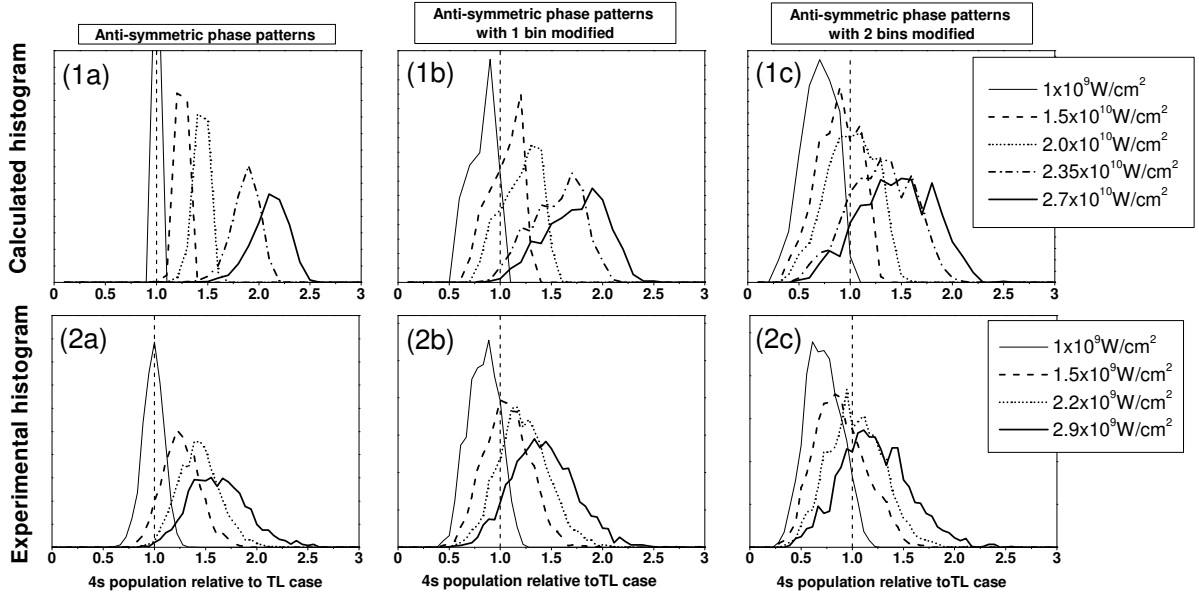}     
\vspace*{-0.7cm} \caption{Histograms for the two-photon absorption
induced in Na by 2500 pulses with different random spectral phase
patterns that are antisymmetric about $\omega_{4s,3s}/2$. Upper
panels: Theoretical fourth-order perturbative results for different
(single-valued) spectral intensities corresponding to different TL
peak intensities $I_{0}$. Lower panels: Experimental
(spatially-integrated) results for different pulse energies
corresponding to different TL peak intensities at the spatial peak
profile. (1a),(2a) -- perfectly antisymmetric phase patterns;
(1b),(2b), and (1c),(2c) -- antisymmetric phase patterns with
reduced antisymmetric nature: each phase pattern has one and two
modified bins respectively. While the two-photon absorption
associated with these phase patterns is lower than with the TL pulse
in the weak-field regime, at sufficiently high intensities it
excells the TL values.
} \label{fig_2}
\end{figure}

First, Figs.\ref{fig_2}(1a) and \ref{fig_2}(2a) present, respectively,
theoretical and experimental enhancement-factor histograms
for the set of 2500 shaped pulses having perfectly antisymmetric phase patterns.
The theoretical results have been calculated numerically using
Eqs.~(\ref{eq4:time-amp})-(\ref{eq10:amp-4th-order}), with each
histogram corresponding to a different (single-valued) intensity $I_{0}$.
Each of the spatially-integrated experimental histograms corresponds
to a different pulse energy, corresponding to the indicated TL peak
intensity at the peak of the spatial beam profile.
The weak-field histograms are located around the enhancement value of one.
As the intensity of the pulse increases, the histogram shifts to
higher enhancement values and gets broader.
The experimental histograms are generally broader than the
theoretical ones due to the integration over the spatial beam
profile and due to the experimental noise. The experimental
signal-to-noise ratio is directly reflected in the width of the
weak-field histogram [Fig.~\ref{fig_2}(2a), thin solid line].

Next, theoretical and experimental absorption results are presented
for phase patterns that are not perfectly antisymmetric around $\omega_{4s,3s}$/2.
For this purpose, in each of the random antisymmetric phase patterns
several of the central 12 bins (out of total 24 bins) 
have been modified to be equal to their bin-pair partner,
such that $\Phi(\omega_{fg}/2-\xi)=\Phi(\omega_{fg}/2+\xi)$.
Figs.~\ref{fig_2}(1b),(2b) and Figs.~\ref{fig_2}(1c),(2c)
show the results for the almost-antisymmetric phase patterns with,
respectively, one and two bins that have been modified.
As compared to the perfectly-antisymmetric set,
the weak-field histograms for these cases are broader and located below the TL value,
i.e., below the enhancement factor of one.
As the pulse intensity increases and deviates from the weak-field regime,
the histogram shifts to higher values and
an increased fraction of the pulses exceeds the TL absorption.
In other words, at high enough intensity the absorption enhancement
beyond the TL absorption is also achieved by phase patterns that
deviate from perfect antisymmetry.

For the perfectly antisymmetric patterns [Fig.~\ref{fig_2}(1a) and (2a)],
the histograms peak for the maximal evaluated intensity
appear at the enhancement factor of 2.2 and 1.6 for the theoretical
and experimental results, respectively.
At the corresponding intensity of 2.7$\times$10$^{10}$W/cm$^{2}$,
all the numerically evaluated phase patterns enhance the two-photon
absorption above the factor of 1.5.
For the experimental results, with a TL peak intensity  of
2.9$\times$10$^{10}$W/cm$^2$ at the peak of the spatial beam
profile, 98$\%$ of the patterns exceed the TL pulse.
For the case of one modified bin [Fig.~\ref{fig_2}(1b) and (2b)],
the weak-field histograms are peaked near the value of 0.9,
i.e., below the TL level, for both calculated and measured results.
As the intensity increases, 99.5$\%$ of the calculated and 93$\%$ of
the measured phase patterns induce an absorption level above the TL
level.
The histograms are located, respectively, around the enhancement
factors of 1.9 and 1.4.
For the case of two modified bins [Fig.~\ref{fig_2}(1c) and (2c)],
the weak-field histograms are located around the value of 0.7.
As the pulse intensity increases, eventually 86$\%$ of the
calculated and 75$\%$ of the measured phase patterns induce an
absorption level above the TL level.
The corresponding resulted histograms are located around the values
of 1.5 and 1.2 respectively.

An example of a typical phase pattern leading to a high enhancement factor
is presented in Fig.\ref{fig_3}.
The corresponding calculated two-photon absorption for this perfectly antisymmetric pattern case
at the intensity of 2.7$\times$10$^{10}$ W/cm$^2$ exceeds the TL absorption by the factor of 3,
corresponding to final $4s$ population of 15$\%$.
The population value is obtained by the exact solution of the
time-dependent Sr\"{o}dinger equation using fourth-order Runge-Kutta
propagation method \cite{amitay_2ph_inter_field2}.
With one and two modified bins, this phase pattern leads to enhancement factors of 2 and 1.35, respectively.
As can be seen, there are clear phase jumps around the bins corresponding to the
resonant transition frequencies of the excitation, $\omega_{4s,3s}$ and $\omega_{7p,4s}$.
Previously, step-like phase patterns have  been shown to lead to the
enhancement of resonance-mediated multiphoton absorption in the
weak-field regime \cite{silberberg_2ph_1plus1,amitay_3ph_2plus1}.
The corresponding mechanism is the conversion of interferences among
near-resonant pathways from a destructive nature (for the TL pulse)
into a constructive one (for the shaped pulse with the phase step).

\section{Discussion}


In the discussion below we analyze the effect of the phase patterns
that enhance the two-photon absorption in the intermediate-field
regime using the perturbative description of
Eq.(\ref{eq4:time-amp})-(\ref{eq10:amp-4th-order}).
We analyze first the functional dependence of the parameterized
amplitudes $A^{(2)}$ and $A^{(R)}$ on the detuning $\delta$ from the
two-photon transition frequency $\omega_{fg}$, which reflects the
interference between the two-photon pathways of the corresponding
type, and as it is shown below has a crucial effect on the
two-photon absorption.
Then we discuss the product of these amplitudes as it contributes to
the fourth-order perturbative term. The aim of this analysis is to
clarify the interference mechanism leading to the enhancement of the
two-photon absorption for the perfectly antisymmetric and modified
phase patterns as compared to the TL pulse.

The amplitudes $A^{(2)}$ and $A^{(R)}$ are shown on Fig.\ref{fig_4}
as a functions of the normalized detuning $\delta/\Delta\omega$,
with $\Delta\omega$ being the pulse spectral bandwidth.
The real and imaginary parts of the complex amplitudes are shown
separately on the different panels: panels Fig.\ref{fig_4} 1(a) and
1(b) for the real and imaginary part of $A^{(2)}$ respectively.
The Raman amplitude $A^{(R)}$ is shown on Fig.\ref{fig_4} 2(a),
2(b), and 3(a), 3(b).
In order to simplify the analysis we discuss first only a
non-resonant part of $A^{(R)}$, i.e. $A^{(R)}=A^{(non-resR)}$
[Fig.\ref{fig_4} 2(a) and 2(b)].
Next, we consider both the non-resonant and resonant parts:
$A^{(R)}=A^{(non-resR)}+A^{(resR)}$ [Fig.\ref{fig_4} 3(a) and 3(b)].
Due to the $\wp$-integration with $1/\delta$ weighting factor in the
near-resonant term $A_{f}^{(near-res)}$
[Eq.(\ref{eq7.3:amp-4th-order})], the mostly dominating region is
the region with small values of $\delta/\Delta\omega$. It is marked
schematically on the corresponding panels.

The term $A^{(2)}(\Omega)$ coherently integrates all the
non-resonant two-photon transition amplitudes
$\widetilde{E}(\omega)\widetilde{E}(\omega_{4s,3s}-\delta-\omega)$
contributed by all the possible pairs of photons with a frequency
sum of $\Omega=\omega_{4s,3s}-\delta$.
The TL pulse induces fully constructive interferences among all
these two-photon pathways for any $\Omega$, the corresponding value
of $A_{TL}^{(2)}(\Omega=\omega_{4s,3s}-\delta)$ is actually the
maximal one (positive and real) for any given $\delta$ as it is
drawn on Fig.\ref{fig_4} (1a),(1b).
As can be seen from Eq.(\ref{eq6:expl-amp-2nd-order-w_fg}),
$A_{TL}^{(2)}(\Omega)$ is a self-convolution of the corresponding
spectrum $\left|\widetilde{E}(\omega)\right|$. Thus, with a Gaussian
spectrum around $\omega_{0}$, it is peaked at $\Omega_{peak} = 2
\omega_{0}$, i.e., at $\delta_{peak} = \omega_{4s,3s} - 2
\omega_{0}$.
For $\omega_{0} < \omega_{4s,3s}/2$ ($\lambda_{0}=$780~nm) we obtain
$\delta_{peak} > 0$ and $A^{(2)}(\omega_{4s,3s}-\delta)$
monotonically increases around $\delta=0$ upon the
negative-to-positive increase of $\delta$, i.e.,
$A^{(2)}(\omega_{4s,3s}-\left|\delta\right|) <
A^{(2)}(\omega_{4s,3s}+\left|\delta\right|)$ for small
$\left|\delta\right|$;

The non-resonant Raman term $A^{(non-resR)}$ coherently integrates
all the non-resonant Raman amplitudes
$\widetilde{E}(\omega+\delta)\widetilde{E}^{*}(\omega)$ contributed
by all the possible photon pairs with frequency difference of
$\delta$.
It is generally symmetric around $\delta=0$, and obtains at that
point its maximal value which depends only on the spectral amplitude
$\left|\widetilde{E}(\omega)\right|$.
For the case of the TL pulse, the amplitude contributed by any pair
of photons is real and positive and so is the resulting total
amplitude $A^{(non-resR)}_{TL}$ [see Fig.\ref{fig_4} (2a),(2b)].
The resonant Raman term $A^{(resR)}$ integrates all the on-resonant
Raman transition amplitudes in its first term as well as
near-resonant Raman transition amplitudes in the second term.
The amplitude corresponding to the photon pathways of the first type
originates from the pairs of photons that are detuned by $\delta$
from the $\left|f\right>$-state but are on-resonant with
$\left|n_{r}\right>$-state ($\delta'=0$) for any possible $\delta$:
$\widetilde{E}(\omega_{fn_{r}}+\delta)\widetilde{E}^{*}(\omega_{fn_{r}})$.
Following the Cauchy principle value theorem, for the TL pulse the
corresponding amplitude contributes to the imaginary part of
$A^{(R)}$ as given in Eq.(\ref{eq10:amp-4th-order}).
The second term integrates the amplitudes contributed by the photon
pairs that are detuned by $\delta'\ne0$ from the
$\left|n_{r}\right>$-state:
$\widetilde{E}(\omega_{fn_{r}}+\delta-\delta')\widetilde{E}^{*}(\omega_{fn_{r}}-\delta')$.
These amplitudes are integrated with the weighting factor
$1/\delta'$ for any possible detuning $\delta'$ and for the TL pulse
contribute to the real part of $A^{(R)}$.
Overall for the TL pulse, $A^{(R)}_{TL}$ has smooth Gaussian shape
[Fig.\ref{fig_4} panels (3a),(3b)] which center is determined by the
spectral amplitude with respect to the three relevant frequencies:
$\omega_{fg}/2$, $\omega_{fn_{r}}$, and $\omega_{0}$, which in our
experiment correspond respectively to 777nm, 781.2nm, and 780nm.

For a phase pattern that is perfectly antisymmetric around
$\omega_{4s,3s}$, the value of
$A^{(2)}_{antisym}(\Omega=\omega_{4s,3s})$ is equal to the
corresponding TL value $A_{TL}^{(2)}(\Omega=\omega_{4s,3s})$ due to
the fully constructive interference among all the involved
two-photon pathways. Similar to the TL pulse, the phases associated
with these pathways are all zero and
$A^{(2)}_{antisym}(\Omega=\omega_{4s,3s})$ has real positive value.
However, as $\Omega$ deviates from $\omega_{4s,3s}$, the real part
of $A^{(2)}(\Omega)$ gradually reduces with comparable magnitude for
positive and negative deviations.
As it appears on Fig.\ref{fig_4} (1a),(1b), the real part of
$A^{(2)}_{antisym}(\omega_{4s,3s}-\delta)$ has a peak at $\delta=0$,
and $\Re
\left\{A^{(2)}_{antisym}(\omega_{4s,3s}-\left|\delta\right|)\right\}
\approx
\Re\left\{A^{(2)}_{antisym}(\omega_{4s,3s}+\left|\delta\right|)\right\}$
for small $\left|\delta\right|$.
Imaginary part of $A^{(2)}_{antisym}(\Omega)$ near $\delta=0$ has a
smooth shape, rising from negative to positive values for the change
from negative to positive detuning and obtains a zero value at
$\delta=0$.

The real part of $A^{(non-resR)}_{antisym}$ [see Fig.~\ref{fig_4},
(2a), (2b)] is symmetrical around $\delta=0$, i.e.
$\Re\left\{A^{(non-resR)}_{antisym}(\omega_{4s,3s}-\left|\delta\right|)
\right\} =
\Re\left\{A^{(non-resR)}_{antisym}(\omega_{4s,3s}+\left|\delta\right|)\right\}$
and obtains its maximal value at $\delta=0$, while the imaginary
part is antisymmetric about $\delta=0$:
$\Im\left\{A^{(non-resR)}_{antisym}(\omega_{4s,3s}-\left|\delta\right|)
\right\} = -
\Im\left\{A^{(non-resR)}_{antisym}(\omega_{4s,3s}+\left|\delta\right|)\right\}$
with $\Im\left\{A^{(non-resR)}_{antisym}(\omega_{4s,3s})\right\}=0$.
The total Raman term corresponding to perfectly antisymmetric phase
pattern $A^{(R)}_{antisym}$ in general is not symmetric. However,
the addition of the $A^{(resR)}$ part approximately conserves the
symmetry of the real part of $A^{(non-resR)}$ near $\delta=0$.
The imaginary part of $A^{(resR)}_{antisym}$ added to the
antisymmetric $\Im\left\{ A^{(non-resR)}_{antisym}\right\}$ gives
approximate symmetry of the total Raman term $A^{(R)}_{antisym}$.
Both real and imaginary parts of $A^{(R)}_{antisym}$ have a peak at
$\delta=0$ and $A^{(R)}_{antisym}(\omega_{4s,3s})=
A^{(R)}_{TL}(\omega_{4s,3s})$.
For the small enough values of the detuning $\delta$, which are of
major importance, we can state that
$A^{(R)}_{antisym}(\omega_{4s,3s}-\left|\delta\right|)\approx
A^{(R)}_{antisym}(\omega_{4s,3s}+\left|\delta\right|)$.

As the individual contributions of the two-photon amplitudes arising
from different types of pathways are described, for the sake of
simplicity of the further discussion, we define a product of the
two-photon and Raman amplitudes $AA(\delta) \equiv
A^{(2)}(\omega_{4s,3s}-\delta)A^{(R)}(\delta)$.
The fourth-order amplitude [Eq.(\ref{eq7.1:amp-4th-order})] is
rewritten and given by
\begin{equation}\label{eq.:4th-order-AA}
    A_{f}^{(4)}=-\frac{1}{i\hbar^{4}}\left|E_{0}\right|^{4}\left[i\pi
    AA(0)-\wp\int_{-\infty}^{+\infty}d\delta\frac{1}{\delta}AA(\delta)\right]
\end{equation}
In the discussion below we consider the cases of non-resonant and of
non-resonant and resonant Raman transitions independently:
$AA^{(nR)}(\delta)$ corresponds to $A^{(R)}=A^{(non-resR)}$
[Fig.~\ref{fig_5} (1a),(1b)] and $AA^{(R)}(\delta)$ corresponds
respectively to $A^{(R)}=A^{(non-resR)}+A^{(resR)}$
[Fig.~\ref{fig_5} (2a),(2b)].
Relying on the analysis above we can summarize the properties of the
products $AA(\delta)$. For the non-resonant Raman transitions only:
\begin{eqnarray}\nonumber
    AA^{(nR)}_{TL}(0)&=&
    AA^{(nR)}_{antisym}(0)\\
    \nonumber
    \Re\left\{AA^{(nR)}_{antisym}(-\left|\delta\right|)\right\}
    &\approx&
    \Re\left\{AA^{(nR)}_{antisym}(+\left|\delta\right|)\right\}\\
    \nonumber
    \Im\left\{AA^{(nR)}_{antisym}(-\left|\delta\right|)\right\}
    &\approx&
    -\Im\left\{AA^{(nR)}_{antisym}(+\left|\delta\right|)\right\}
    \nonumber
\end{eqnarray}
and for both non-resonant and resonant Raman transitions:
\begin{eqnarray}\nonumber
    AA^{(R)}_{TL}(0)&=&
    AA^{(R)}_{antisym}(0)\\
    \nonumber
    \Re\left\{AA^{(R)}_{antisym}(-\left|\delta\right|)\right\}
    &\approx&
    \Re\left\{AA^{(R)}_{antisym}(+\left|\delta\right|)\right\}\\
    \nonumber
    \Im\left\{AA^{(R)}_{antisym}(-\left|\delta\right|)\right\}
    &\approx&
    \Im\left\{AA^{(R)}_{antisym}(+\left|\delta\right|)\right\}
    \nonumber
\end{eqnarray}
These properties suggest that for the antisymmetric phase patterns,
approximate symmetry of $AA^{(R)}_{antisym}$ near $\delta=0$ leads
to sufficient suppression of the principle value integral in
Eq.(\ref{eq.:4th-order-AA}) as the integrand factor $1/\delta$ is
antisymmetric about $\delta=0$.
On the other hand, for the TL pulse $AA^{(R)}_{TL}$ is not symmetric
and the value of the corresponding integral is not as small as for
the antisymmetric phase pattern:
\begin{eqnarray}\label{eq.:AA-nr}
    \left|\int_{-\infty}^{+\infty}d \delta\frac{1}{\delta}
    AA^{(nR)}_{TL}(\delta)\right|
    >
    \left|\int_{-\infty}^{+\infty}d \delta\frac{1}{\delta}
    AA^{(nR)}_{antisym}(\delta)\right|
\end{eqnarray}
and
\begin{eqnarray}\label{eq.:AA}
    \left|\int_{-\infty}^{+\infty}d \delta\frac{1}{\delta}
    AA^{(R)}_{TL}(\delta)\right|
    >
    \left|\int_{-\infty}^{+\infty}d \delta\frac{1}{\delta}
    AA^{(R)}_{antisym}(\delta)\right|
\end{eqnarray}

The amplitude of the final state $\left|f\right>$ is determined by
the interference between the second and the fourth order amplitudes.
The nature of the interference (constructive or destructive) between
$A^{(2)}_{4s}$ and $A^{(4)}_{4s}$ is set by the pulse spectrum and
by the non-resonant Raman couplings sum
$(\mu_{3s,3s}^{2}+\mu_{4s,4s}^{2})$.
We have determined that for the case of atomic sodium excited with
red detuned pulse this interference is destructive: the fourth order
amplitude $A^{(4)}_{4s}$ suppresses the second order amplitude
$A^{(2)}_{4s}$. Recalling that $A^{(2)}_{4s}\propto I_{0}$ and
$A^{(4)}_{4s}\propto I_{0}^{2}$ we write the value of the
enhancement factor corresponding to the shaped pulse and normalized
by the TL one:
\begin{equation}\label{eq.:R_antisym}
    R_{shaped}=\frac{P_{4s, shaped}}{P_{4s, TL}}=
    \frac{\left|A^{(2)}_{shaped}+A^{(4)}_{shaped}\right|^{2}}
    {\left|A^{(2)}_{TL}+A^{(4)}_{TL}\right|^{2}}=
    \frac{\left|\kappa_{shaped}+{\cal{K}}_{shaped}I_{0}\right|^{2}}
    {\left|1+{\cal{K}}_{TL}I_{0}\right|^{2}}.
\end{equation}
For the antisymmetric phase patterns
$\kappa_{shaped}$=$\kappa_{antisym}$=1 as
$A^{(2)}_{antisym}=A^{(2)}_{TL}$, while from the analysis above it
follows that
$\left|{\cal{K}}_{antisym}\right|<\left|{\cal{K}}_{TL}\right|$, and
${\cal{K}}_{TL},{\cal{K}}_{antisym}<0$, resulting in the value of
$R_{antisym}>1$.
Eq.(\ref{eq.:R_antisym}) reflects the mechanism of the two-photon
absorption enhancement by the antisymmetric phase patterns. As the
pulse intensity $I_{0}$ increases, the relative weight of
$A^{(4)}_{f}$ is increased and the enhancement effect is highly
pronounced.

\begin{figure} 
\includegraphics[width=10cm]{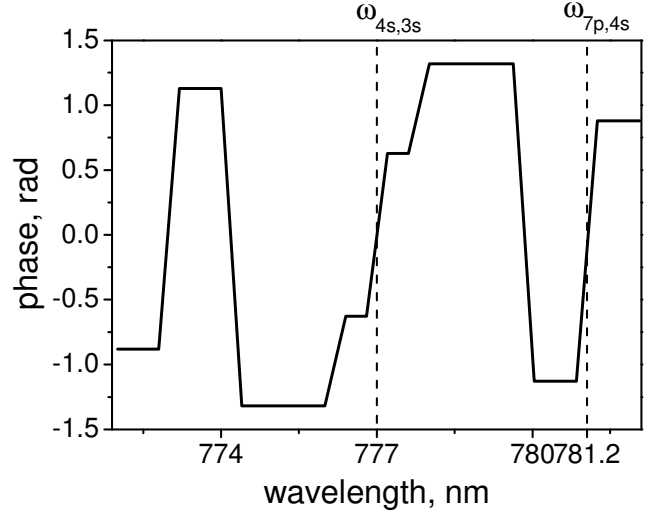}     
\vspace*{-0.7cm}
\caption{
An example of highly-enhancing phase pattern that is antisymmetric about $\omega_{4s,3s}/2$.
The corresponding population transfer induced by the two-photon absorption in the upper limit of the intermediate-field regime
(2.7$\times$10$^{10}$W/cm$^2$)
is calculated to be 15$\%$, which corresponds to an enhancement factor of 3 over the TL absorption.
The phase pattern is characterized by step-like phase changes at the resonant frequencies of the system:
$\omega_{4s,3s}/2$ (777~nm) and $\omega_{7p,4s}$ (781.2~nm).
See the text for a detailed description.
} \label{fig_3}
\end{figure}

\begin{figure} 
\includegraphics[width=17cm]{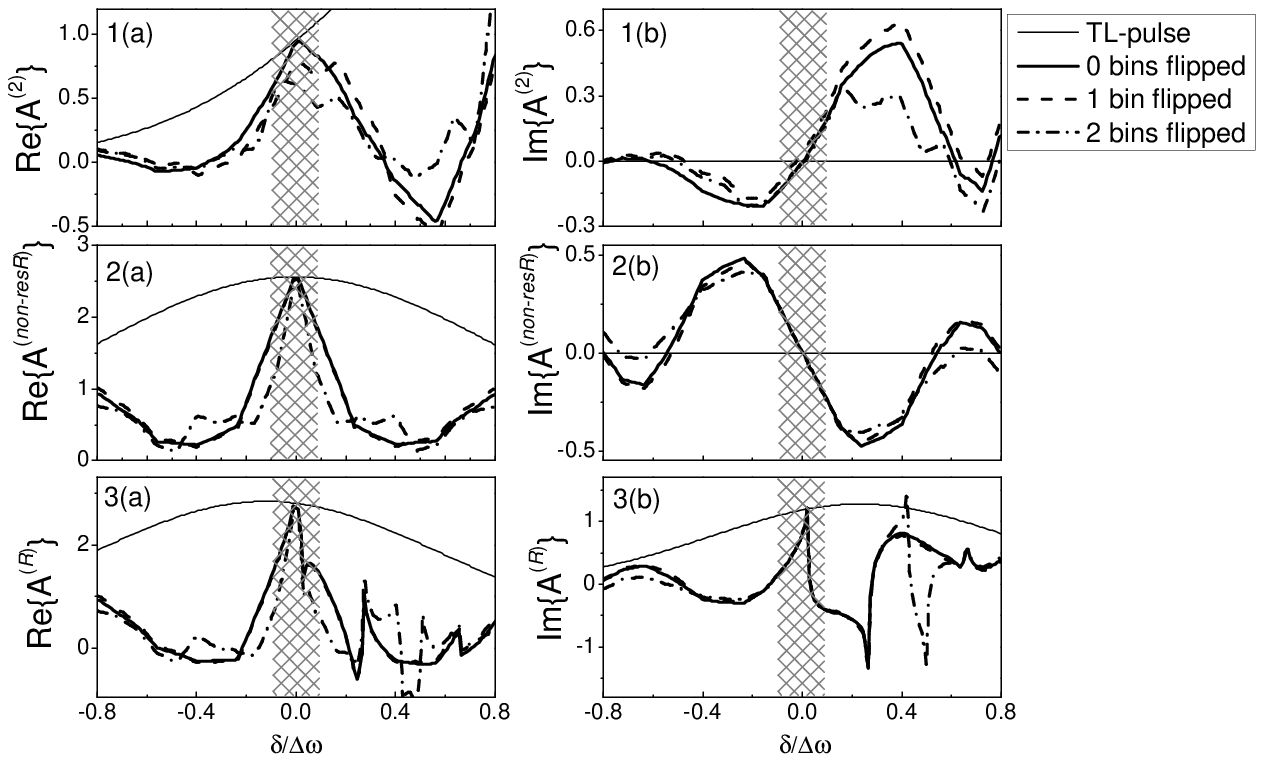}     
\vspace*{-0.7cm} \caption{Theoretical results for parametrized
amplitudes $A^{(2)}(\delta)$, $A^{(non-resR)}(\delta)$, and
$A^{(R)}(\delta)$ that are introduced by the frequency-domain
description and used for explanation of the intermediate-field
behavior of the shaped pulses. The results for the four pulse shapes
are presented: TL pulse, perfectly antisymmetric phase pattern given
on Fig.~\ref{fig_3}, and almost antisymmetric phase patterns with
one and two modified bins. The amplitudes are represented as a
function of normalized detuning $\delta/\Delta\omega$ [see text]
with the most contributing region of small $\left|\delta\right|$
indicated schematically by the dashed area.
} \label{fig_4}
\end{figure}

\begin{figure} 
\includegraphics[width=17cm]{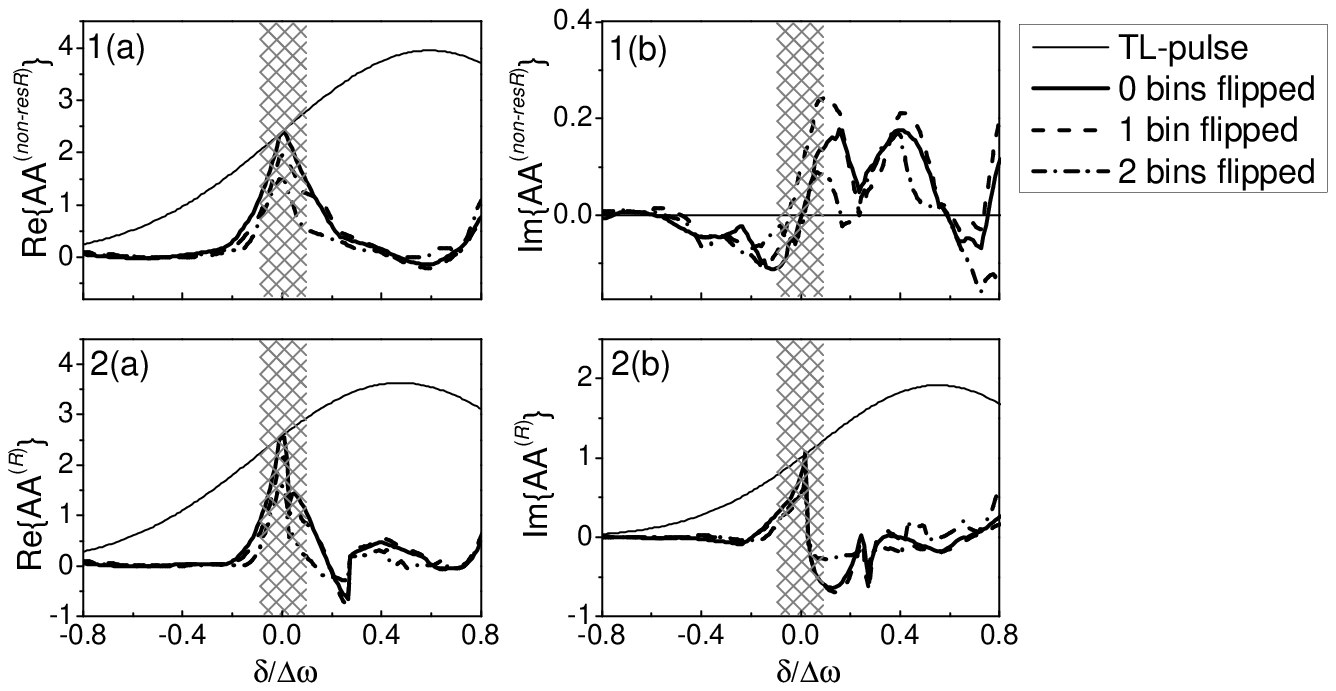}     
\vspace*{-0.7cm}
\caption{Theoretical results for parametrized
amplitudes $AA^{(non-resR)}(\delta)$ and $AA^{(R)}(\delta)$ that are
introduced by the frequency-domain description and used for
explanation of the intermediate-field behavior of the shaped pulses.
The results for the four pulse shapes are presented: TL pulse,
perfectly antisymmetric phase pattern given on Fig.~\ref{fig_3}, and
almost antisymmetric phase patterns with one and two modified bins.
The amplitudes are represented as a function of normalized detuning
$\delta/\Delta\omega$ [see text] with the most contributing region
of small $\left|\delta\right|$ indicated schematically by the dashed
area.
} \label{fig_5}
\end{figure}

Consider now the antisymmetric phase patterns with several modified
bins. For the representative phase pattern [Fig.~\ref{fig_3}], the
cases of one and two flipped bins are considered, and
$A^{(2)}(\delta)$, $A^{(R)}(\delta)$ and $AA(\delta)$ are drawn on
Figs.~\ref{fig_4} and \ref{fig_5} respectively.
The interference of the pathways corresponding to the flipped bins
with the rest of the pathways has turned to be a destructive one.
This fact is reflected in the value of $A^{(2)}_{4s}$ for these
phase patterns that is reduced and does not reach the corresponding
value of the TL pulse:
$A^{(2)}_{4s,2bins}<A^{(2)}_{4s,1bin}<A^{(2)}_{4s,antisym}=A^{(2)}_{4s,TL}$.
The peak shape of $A^{(2)}(\delta)$ has also been changed as
compared to the antisymmetric phase pattern.
On the other hand, although the peak shape of $A^{(non-resR)}$ has
been changed, it remains symmetrical about $\delta=0$ and the value
of $A^{(non-resR)}(0)$ remains equal to that of the TL pulse as it
is actually phase-pattern-independent.
Similar observations are made concerning the $A^{(R)}$ term:
$A^{(R)}_{2bins}(0)=A^{(R)}_{1bin}(0)=A^{(R)}_{antisym}(0)=A^{(R)}_{TL}(0)$,
while the shape of $A^{(R)}(\delta)$ has been changed.
Overall, the product of the amplitudes $AA(\delta)$ incorporate the
properties of its both components:
$AA_{2bins}(0)<AA_{1bin}(0)<AA_{antisym}(0)=AA_{TL}(0)$, while the
shape of $AA(\delta)$ remains close to symmetrical to a good
approximation at the mostly contributing region of small values of
$\left|\delta\right|$.
Consequently, the contribution of the on-resonant pathways
($\delta=0$) to the fourth-order amplitude is less for the modified
antisymmetric phase patterns than that of the perfectly
antisymmetric ones, but the contribution of the near-resonant
pathways ($\delta\ne0$) has not been changed sufficiently, as the
symmetry of the Cauchy's principle value integrand is approximately
conserved.

We analyze now the enhancement factor $R$ (Eq.\ref{eq.:R_antisym})
considering the phase patterns with the flipped bins. From the
second perturbative order term contribution we obtain
$\kappa_{2bins}<\kappa_{1bin}<\kappa_{antisym}=\kappa_{TL}=1$, while
from the fourth order one:
$\left|{\cal{K}}_{2bins}\right|<\left|{\cal{K}}_{1bin}\right|<\left|{\cal{K}}_{antisym}\right|<\left|{\cal{K}}_{TL}\right|$.
These results affect the values of the total amplitude $A_{4s}$
corresponding to the weak-field limit [$A_{4s}=A^{(2)}_{4s}$] to be
$A_{4s,2bins}<A_{4s,1bin}<A_{4s,antisym}=A_{4s,TL}$. The
corresponding normalized R-values in the weak-field regime are:
$R_{antisym}$=1, $R_{1bin}$=0.7, and $R_{2bins}$=0.4.
However, at the higher intensities, the intensity-dependent term
${\cal{K}}I_{0}$ plays a significant role,  increasing the value of
the enhancement factor $R$.
The mechanism of the enhancement follows from the interference
nature between the perturbative orders: although the values of
$A^{(2)}_{4s}$ are smaller for the modified antisymmetric phase
pattern, the attenuation by $A^{(4)}_{4s}$ stays low, close to the
values of the perfectly antisymmetric pattern, as compared to the TL
pulse.
At sufficiently high intensity the resulted total population of the
final state $P_{4s}=\left|A_{4s}\right|^2$ exceed the population
corresponding to the TL pulse.
At the upper intermediate-field regime we obtain for the phase
pattern on Fig.~\ref{fig_3} that while its enhancement factor is
$R_{antisym}$=3, for the modified phase patterns $R_{1bin}$=2, and
$R_{2bins}$=1.35.

\section{Conclusions}

In conclusion, we have identified an extended family of phase
shaped-pulses that, as opposed to the weak-field regime, in the
intermediate-field regime enhance the two-photon absorption beyond
the absorption level induced by the transform-limited pulse. This
corresponding phase patterns are of antisymmetric nature about
one-half the two-photon transition frequency.
The phase-patterns are demonstrated to be robust to imperfections in their antisymmetry:
even though the largest enhancement is achieved with phase patterns of perfect antisymmetry,
at high enough intermediate-field intensities absorption enhancement 
is still achievable even with 
patterns that deviate from perfect antisymmetry. 
The degree of tolerance to deviations from perfect antisymmetry increases as the pulse intensity increases.
These findings are of 
importance for the development of rational selective coherent control strategies,
for reducing the search space when automatic feedback control optimization strategies are employed
\cite{rabitz_feedback_learning_idea,gerber_feedback_control_review},
and for nonlinear spectroscopy and microscopy with shaped femtosecond pulses. 
A particularly interesting scenarios is when the excitation simultaneously involves multiple excitation channels \cite{amitay_multi-channel}.
Then, the tolerance to deviation from perfect antisymmetry
allows to keep the excitation of the two-photon channel sufficiently high,
while adjusting the excitation of the other channels.


\section*{ACKNOWLEDGMENTS}
This research was supported by The Israel Science Foundation (grant
No. 127/02), by The James Franck Program in Laser Matter
Interaction, by The Technion President Fund,
and by The Technion's Fund for The Promotion of Research.



\newpage

\end{document}